\documentclass[sigconf]{acmart}
 \renewcommand\footnotetextcopyrightpermission[1]{} 
 \pdfoutput=1
\usepackage{tikz}
\usepackage{multirow}
\usepackage{amsmath}
\usepackage{qcircuit}
\usepackage{algorithm}
\usepackage{algpseudocode}
\usepackage{graphicx}
\usepackage{caption}
\usepackage{standalone}
\usepackage{subcaption}
\usepackage{float}

\usepackage{tikz}
\usepackage{multirow}
\usepackage{amsmath}
\usepackage{qcircuit}
\usepackage{algorithm}
\usepackage{algpseudocode}
\usepackage{graphicx}
\usepackage{caption}
\usepackage{standalone}
\usepackage{subcaption}
\usepackage{float}

\newcommand{\name}{Coqa}

\usetikzlibrary{shapes,arrows,positioning}
\usetikzlibrary{shadows,arrows.meta,positioning,backgrounds,fit}

\usetikzlibrary{external}
\begin{document}

\title{Coqa: Blazing Fast Compiler Optimizations for QAOA}
\author{Yuchen Zhu*}
\affiliation{%
  \institution{Rensselaer Polytechnic Institute}
  \city{Troy}
  \state{NY}
  \country{USA}
}

\author{Yidong Zhou*}
\affiliation{%
  \institution{Rensselaer Polytechnic Institute}
  \city{Troy}
  \state{NY}
  \country{USA}}
  
\author{Jinglei Cheng*}
\affiliation{%
  \institution{Purdue University}
  \city{West Lafayette}
  \state{IN}
  \country{USA}
}

\author{Yuwei Jin*}
\affiliation{%
  \institution{Rutgers University}
  \city{New Brunswick}
  \state{NJ}
  \country{USA}}
  
\author{Boxi Li}
\affiliation{%
  \institution{Forschungszentrum Jülich
}
  \city{Jülich}
  \country{Germany}}
  
\author{Siyuan Niu}
\affiliation{%
 \institution{Lawrence Berkeley National Laboratory}
 \city{Berkeley}
 \state{CA}
 \country{USA}}

\author{Zhiding Liang}
\affiliation{%
  \institution{Rensselaer Polytechnic Institute}
  \city{Troy}
  \state{NY}
  \country{USA}}

\thanks{*These authors contributed to the work equally.}

\begin{abstract}
The Quantum Approximate Optimization Algorithm (QAOA) is one of the most promising candidates for achieving quantum advantage over classical computers. 
However, existing compilers lack specialized methods for optimizing QAOA circuits. 
There are circuit patterns inside the QAOA circuits, and current quantum hardware has specific qubit connectivity topologies. 
Therefore, we propose~\name~to optimize QAOA circuit compilation tailored to different types of quantum hardware. 
Our method integrates a linear nearest-neighbor (LNN) topology and efficiently map the patterns of QAOA circuits to the LNN topology by heuristically checking the interaction based on the weight of problem Hamiltonian. 
This approach allows us to reduce the number of SWAP gates during compilation, which directly impacts the circuit depth and overall fidelity of the quantum computation.
By leveraging the inherent patterns in QAOA circuits, our approach achieves more efficient compilation compared to general-purpose compilers.
With our proposed method, we are able to achieve an average of 30\% reduction in gate count and a 39x acceleration in compilation time across our benchmarks.

\end{abstract}

\maketitle
\pagestyle{plain}

\section{Introduction}

Quantum computing is an emerging technology with the potential to tackle problems that are currently beyond the reach of classical computation, such as tasks in cryptography~\cite{shor1999polynomial}, quantum chemistry~\cite{cao2019quantum,wang2022quantumnas,liang2024napa}, and optimization~\cite{li2020quantum,liang2023hybrid}. 
Quantum computers' ability to process complex computations in parallel, leveraging quantum phenomena like superposition and entanglement, offers great potentials for industries ranging from drug discovery to secure communications. 
However, different types of noise in quantum hardware, which can cause quantum states to collapse and lead to computational errors, presents significant challenges, hindering the realization of practical quantum applications. 
Overcoming these obstacles is essential for unlocking the full potential of quantum computing and making it a viable tool for solving real-world problems.

Quantum error correction (QEC) has been proposed as a solution to mitigate noise and enable fault-tolerant quantum computing (FTQC). 
However, the substantial overhead required—often involving millions of qubits—renders it impractical for the near-term devices we have today.
Currently, we are in the noisy intermediate-scale quantum (NISQ) era~\cite{Preskill2018NISQ}, characterized by quantum systems with a limited number of qubits of constrained quality. 
In this context, variational quantum algorithms (VQAs)~\cite{peruzzo+:nature14, cheng2022topgen, wang2022qoc,liang2022variational, jin2024tetris, zhuang2024improving,phalak2024non} have emerged as a more practical approach. 
These hybrid quantum-classical algorithms leverage the strengths of both quantum and classical computing, requiring fewer qubits and shallower circuit depths. 
They are specifically designed to exploit the potential of NISQ devices, aiming to achieve quantum advantages in the near term.

The Quantum Approximate Optimization Algorithm (QAOA)~\cite{farhi2014quantum, zhu_2023_qaoa, zhang2024groverqaoa3} stands out as one of the most promising algorithms within the realm of VQAs for addressing combinatorial optimization problems, such as Max-Cut~\cite{guerreschi2019qaoa} and Low Autocorrelation Binary Sequence (LABS)~\cite{shaydulin2024evidence}. 
It leverages the power of quantum mechanics to explore complex solution spaces more efficiently than classical algorithms. 
QAOA works by preparing a parameterized quantum circuit that encodes the problem, and then iteratively refining the parameters through a classical-quantum optimization loop to minimize a classical cost function tailored to the specific problem at hand. 
This hybrid approach allows QAOA to potentially find high-quality solutions more quickly than traditional methods, making it a key contender for demonstrating quantum advantage in practical applications.

When evaluating QAOA on quantum hardware, it is essential to compile the quantum circuits to the specific architecture of the hardware. 
For example, superconducting quantum hardware typically has nearest-neighbor connectivity constraints, which create challenges in executing quantum circuits as intended. 
A critical step of this compilation is known as the qubit mapping problem, which involves aligning the quantum circuit with the hardware's constraints.
This process includes two key steps: first, determining the initial layout by mapping the circuit qubits onto the hardware's physical qubits, and second, inserting SWAP gates to ensure that two-qubit gates meet the hardware's topology requirements. 
Our paper focuses on optimizing this mapping process within the compiler for QAOA circuits, aiming to enhance the performance and efficiency of quantum computations on current hardware.

When optimizing the mapping process for quantum circuits, key objectives often include reducing the number of SWAP gates, especially since these gates typically have higher error rates—enhancing the overall circuit fidelity, and minimizing circuit depth to improve execution efficiency. 
Achieving these goals is crucial for maintaining the integrity of quantum computations, particularly in the noisy environment of current quantum hardware. 
While general-purpose compilers like Qiskit~\cite{aleksandrowicz2019qiskit}, TKet~\cite{Sivarajah_TKET_A_Retargetable_2020}, and BQSKit~\cite{osti_1785933} are designed to handle a wide range of quantum circuits, QAOA circuits possess unique properties that call for specialized compilers to achieve optimal performance. 

One of the distinct characteristics of QAOA circuits is that their two-qubit gates are commutative, meaning the order in which these gates are applied can be adjusted without affecting the final outcome. 
This property provides additional flexibility in re-ordering the gates during the optimization process. 
Specialized QAOA compilers can take advantage of this flexibility, allowing for more efficient circuit layouts and reducing the number of required SWAP gates~\cite{jin2021structured, alam2020circuit, alam2020efficient}. 
Moreover, unlike other algorithms that might benefit from the specific connectivity features of IBM's heavy hex hardware, QAOA circuits do not necessarily require such complex connectivity. 
Instead, a linear transformation of the topology map can be employed to achieve lower compilation overhead, accelerating the process and further enhancing circuit efficiency. 
By leveraging these approaches, our QAOA-specific compiler can significantly improve the performance and reliability of quantum computations.


Our contributions are as follows:
\begin{itemize}
    \item We are the first to explore the LNN topology for QAOA circuits.
    \item We analyze the structure of QAOA circuits and the LNN topology, and propose a method to efficiently map QAOA circuits onto the LNN topology.
    \item We evaluate our method on IBM's heavy-hex topology and demonstrate a 30\% reduction in gate count and a 39x acceleration in compilation time.

\end{itemize}

The rest of the paper is structured as follows: In Section~\ref{section_background}, we introduce the background of QAOA and LNN. In Section~\ref{section_motivation}, we present the motivation for our work. Section~\ref{section_methodology} discusses the technical details of our method, and Section~\ref{section_evaluation} demonstrates the results on IBM quantum computers. We cover related work in Section~\ref{section_relatedwork} and conclude in Section~\ref{section_conclusion}.

\label{introduction}

\section{Background}
\label{section_background}
In this session, we explore the fundamental concepts associated with QAOA and the challenges of its implementation on quantum hardware. The primary challenges involve the compilation and optimization of quantum circuits to align with the constraints of modern quantum architectures.

\subsection{Introduction to QAOA}
QAOA is a hybrid quantum-classical algorithm widely used to find approximate solutions to combinatorial optimization problems. 
More specifically, QAOA involves five main steps: problem setup, unitary transformation, ansatz construction, classical optimization, and measurement.

\begin{enumerate}
    \item \textbf{Problem Setup}: Encode the optimization problem into a cost Hamiltonian \( H_C \), with the goal of finding its ground state.
    \item \textbf{Unitary Transformations}: Define the unitary operations:
    \[
    U(H_C, \gamma) = e^{-i \gamma H_C}, \quad U(H_B, \beta) = e^{-i \beta H_B}
    \]
    where \( H_C \) is the cost Hamiltonian, \( H_B \) is a mixing Hamiltonian, and \( \gamma \) and \( \beta \) are parameters to be optimized.
    \item \textbf{Ansatz Construction}: Create the ansatz state \( |\psi(\gamma, \beta)\rangle \) by applying a series of unitary transformations starting from an initial state \( |\psi_0\rangle \):
    \[
    |\psi(\gamma, \beta)\rangle = U(H_B, \beta_p) U(H_C, \gamma_p) \ldots U(H_B, \beta_1) U(H_C, \gamma_1) |\psi_0\rangle
    \]
    \item \textbf{Classical Optimization}: Optimize the parameters \( \gamma \) and \( \beta \) using classical algorithms to minimize the expectation value of \( H_C \):
    \[
    \langle H_C \rangle = \langle \psi(\gamma, \beta) | H_C | \psi(\gamma, \beta) \rangle
    \]
    \item \textbf{Measurement}: Measure the quantum state \( |\psi(\gamma, \beta)\rangle \) to obtain a classical bit string, representing an approximate solution to the optimization problem.
\end{enumerate}

\subsection{Compilation of QAOA}

QAOA involves creating a quantum circuit with alternating layers of unitary transformations and problem-specific Hamiltonian interactions.
The challenge in compiling QAOA circuits stems from their specific structure and the need to align them with hardware constraints.

\subsubsection{Problem-Graph Induced Circuit}
\begin{figure}[b!]
    \begin{subfigure}[t!]{0.17\linewidth}
        \centering
        \includegraphics[width=\linewidth]{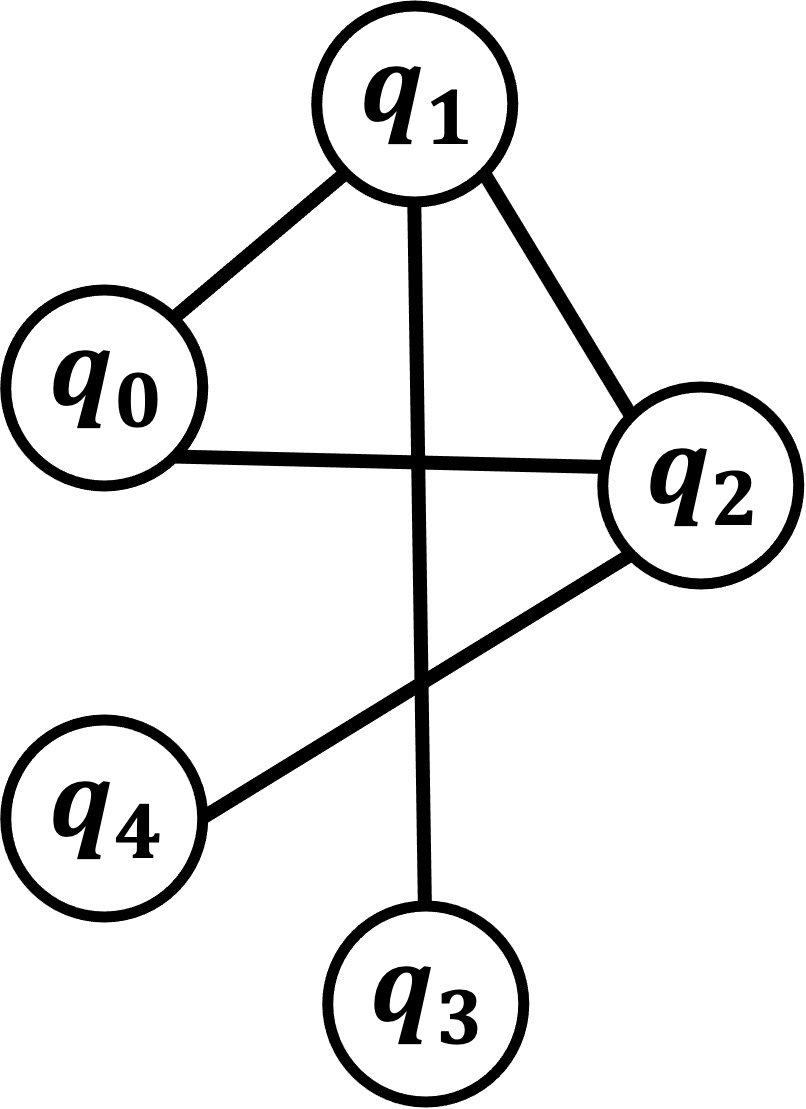}    
        \label{fig:inputgraph}
        \caption[]{} 
    \end{subfigure}
    \hspace{0.015\linewidth}
    \begin{subfigure}[t!]{0.38\linewidth}  
        \centering 
        \includegraphics[width=\linewidth]{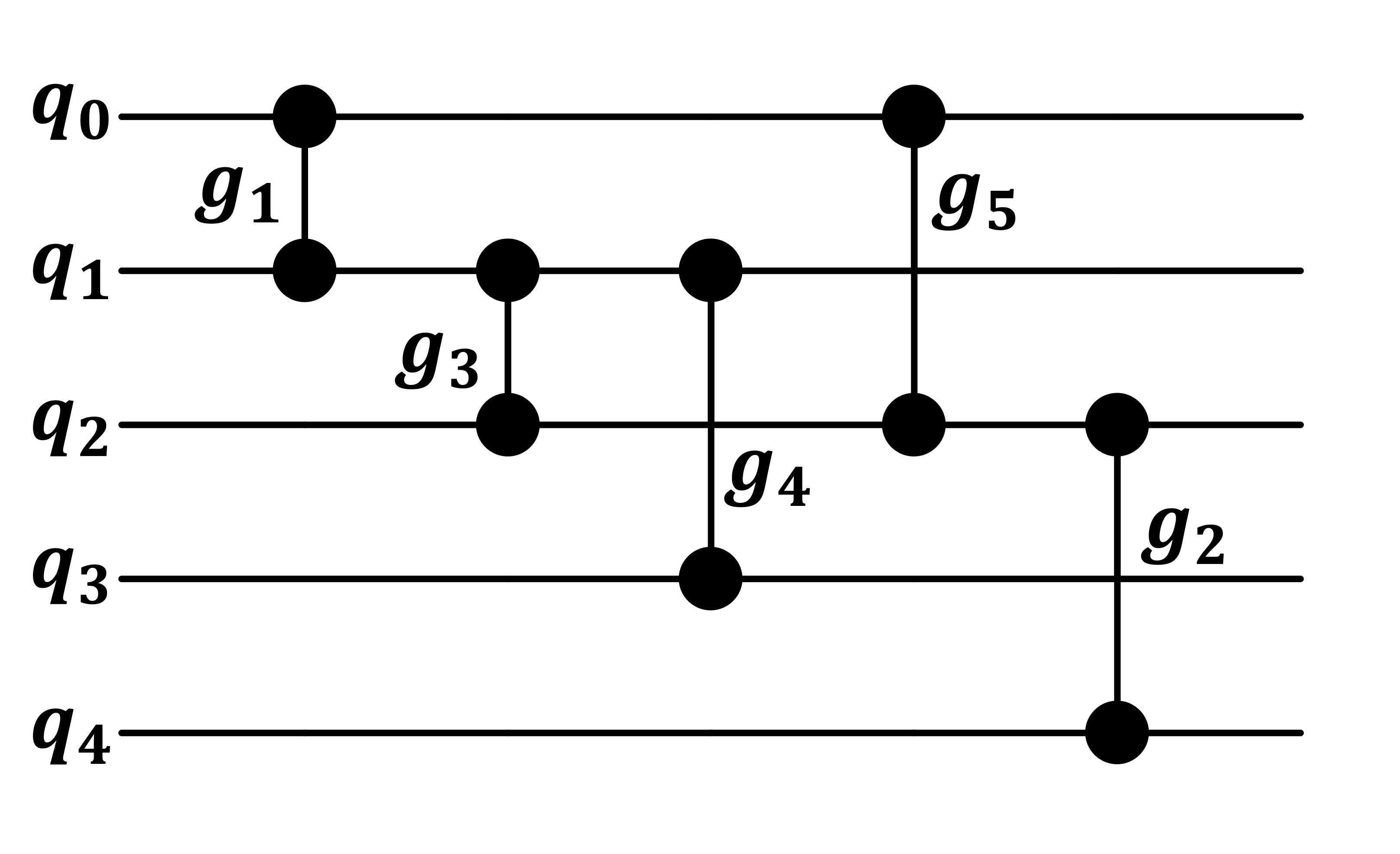}  
        \label{fig:validcircuits1}
        \caption[]{}
    \end{subfigure}
    \hspace{0.015\linewidth}
    \begin{subfigure}[t!]{0.34\linewidth}   
        \centering 
        \includegraphics[width=\linewidth]{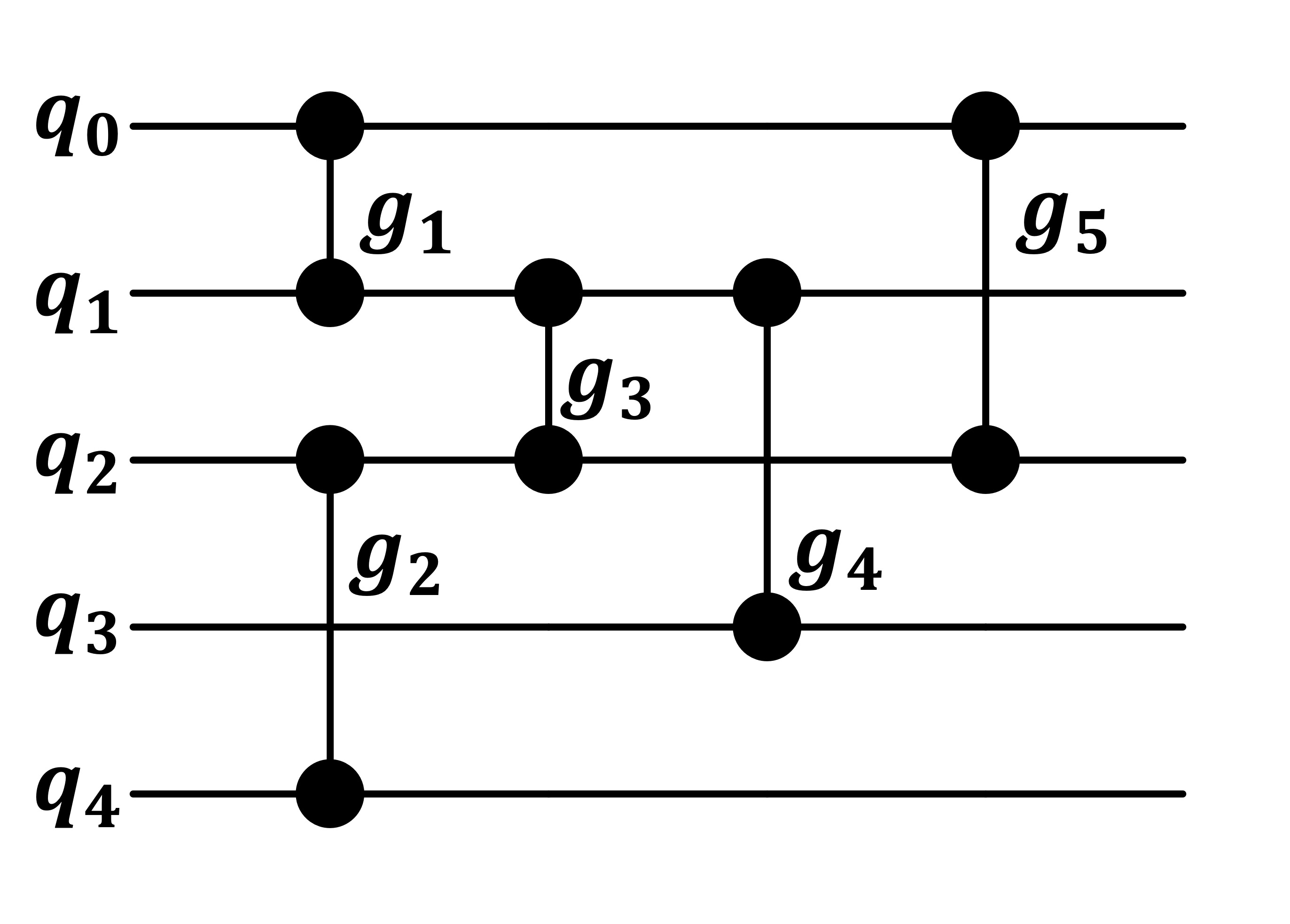}    
        \label{fig:validcircuits2}
        \caption[]{}
    \end{subfigure}
    \caption[]{A five-qubit QAOA-Maxcut circuit example: (a) the input problem graph. (b) and (c) are two valid circuits.} 
    \label{fig:problemgraph}
\end{figure}

The structure of a QAOA circuit is dictated by the input problem graph. For instance, in the QAOA-Maxcut problem, qubits correspond to vertices in the graph, and controlled-phase (CPHASE) gates correspond to edges. This unique feature of QAOA circuits is their flexibility in gate ordering due to the commutativity of two-qubit gates. As illustrated in Fig. \ref{fig:problemgraph}, multiple valid gate arrangements can be applied to the same problem graph, allowing for optimization in circuit compilation.

\subsubsection{Compilation Objectives}
QAOA circuits are characterized by their use of parameterized unitary gates, which are optimized through classical feedback loops.
The compilation process must address the following parts:
\begin{itemize}
    \item \textbf{Qubit Mapping and Hardware Constraints}: A major challenge in compiling QAOA circuits is efficiently mapping logical qubits to physical qubits on a quantum chip, which involves addressing connectivity constraints. Due to limited qubit connectivity and the potential for gate errors, logical qubits often need to be connected using SWAP gates. This mapping must also account for hardware limitations such as gate fidelities and qubit coherence times to ensure accurate computation.
    \item \textbf{Gates Optimization}: Optimizing the order of unitary gates in QAOA circuits is crucial for minimizing circuit depth and error rates. The commutativity of two-qubit gates in QAOA provides flexibility in gate sequencing, which can be leveraged to enhance performance. 
\end{itemize}

\begin{figure}[t!]
    \begin{subfigure}[t!]{0.07\linewidth}
        \centering
        \includegraphics[width=\linewidth]{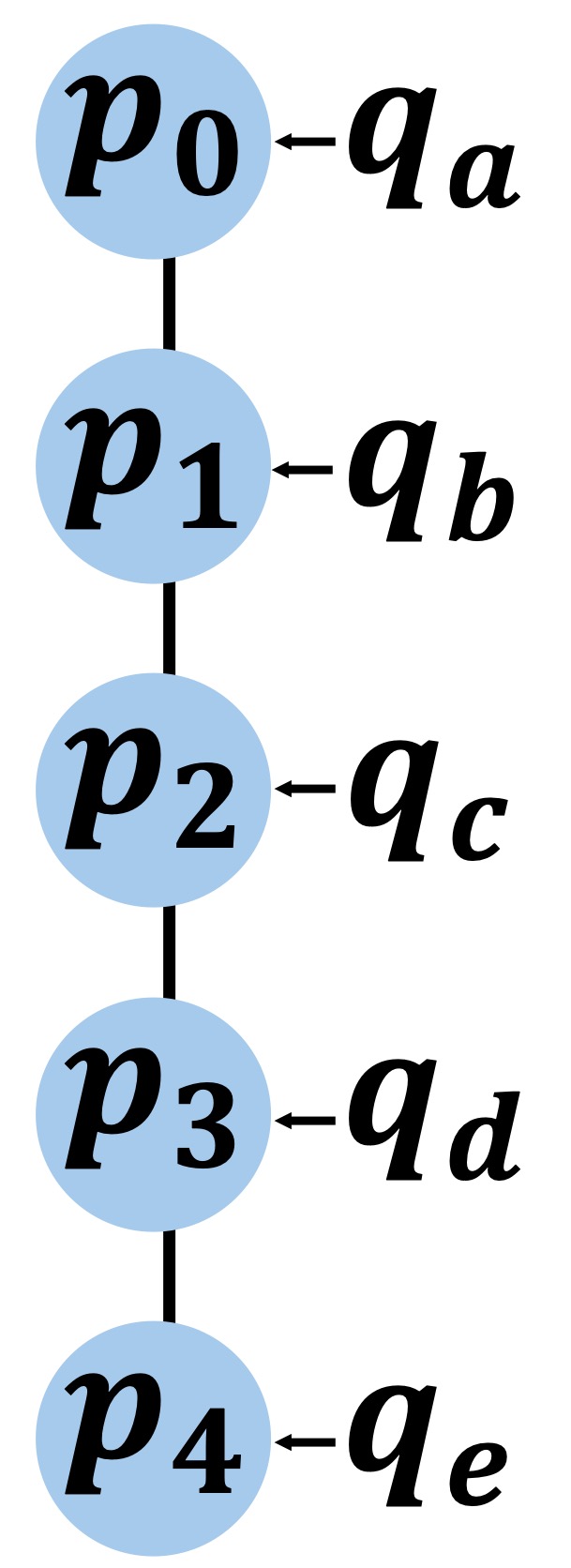}    
        \label{fig:hardware coupling}
        \caption[]{} 
    \end{subfigure}
    \hspace{0.015\linewidth}
    \begin{subfigure}[t!]{0.43\linewidth}  
        \centering 
        \includegraphics[width=\linewidth]{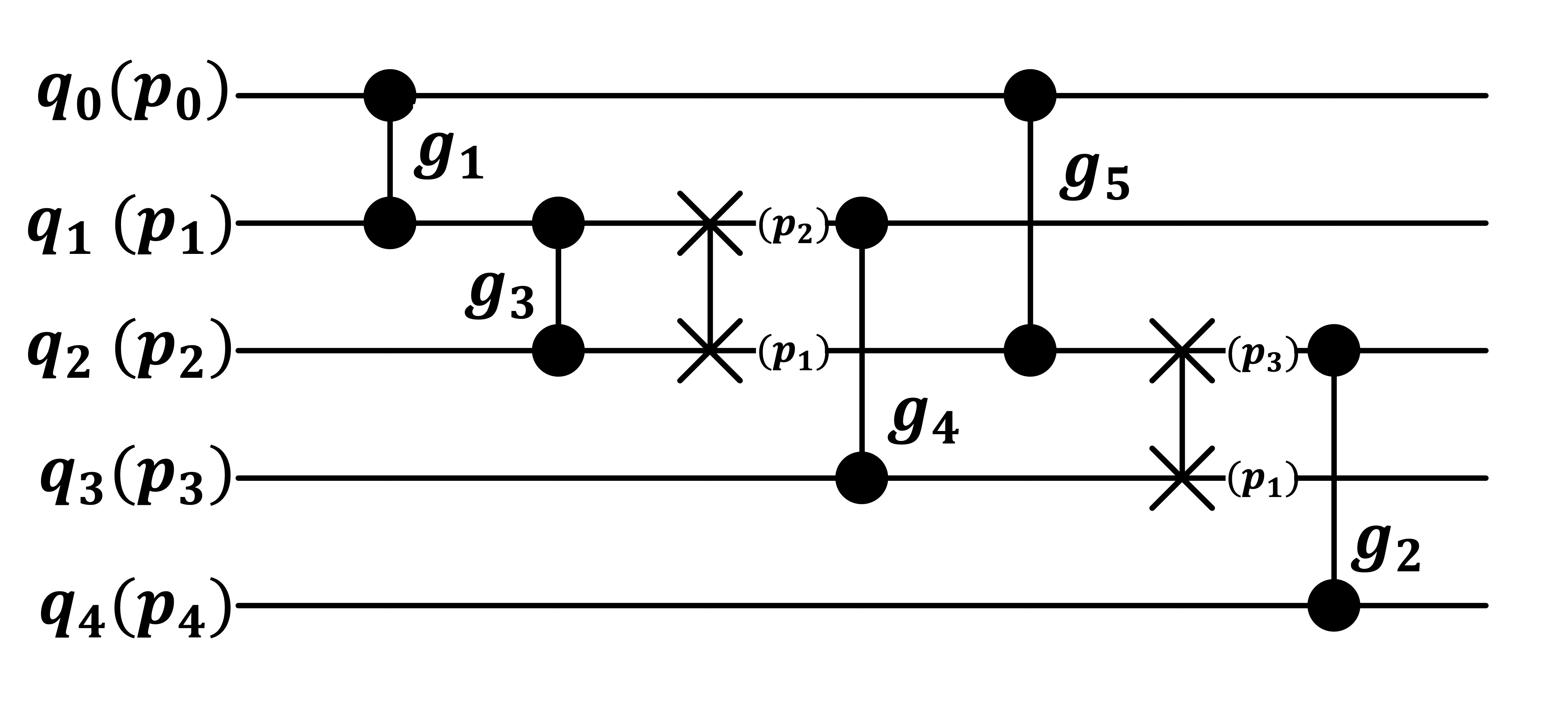}  
        \label{fig:possible compiled circuit1}
        \caption[]{}
    \end{subfigure}
    \hspace{0.015\linewidth}
    \begin{subfigure}[t!]{0.41\linewidth}   
        \centering 
        \includegraphics[width=\linewidth]{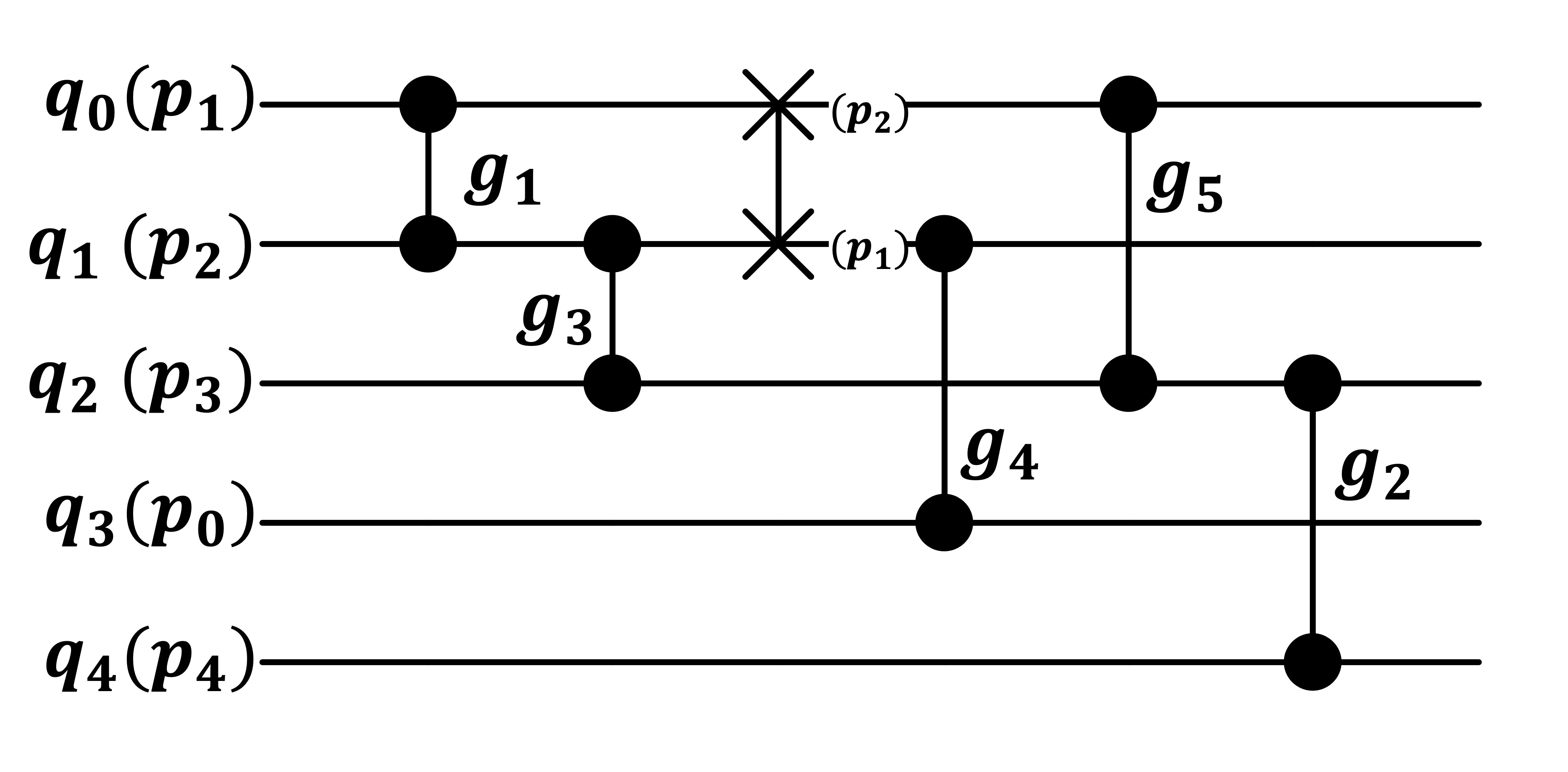}    
        \label{fig:possible compiled circuit2}
        \caption[]{}
    \end{subfigure}
    \caption[]{(a) Hardware coupling map; (b) One possible compiled version of the circuit corresponding to the input problem graph, (c) Another possible compiled version of the same circuit. Due to the qubit mapping process, the number of gates and the depth of the circuit in (c) are reduced compared to (b), illustrating a more efficient compilation.} 
    \label{fig:qubitmapping}
\end{figure}

By addressing these aspects—problem-graph structure, qubit mapping, and gate optimization—the compilation of QAOA circuits aims to enhance performance and adapt to the constraints of quantum hardware effectively. Fig. \ref{fig:qubitmapping} illustrates how different optimization strategies can affect circuit depth and gate count, highlighting the impact of gate ordering and SWAP insertion on overall circuit efficiency.

\section{Motivation}
\label{section_motivation}

\paragraph{\textbf{Limitation in Quantum Hardwares.}} 
Quantum devices are inherently error-prone, and the variability in error rates across different qubits and gates can have a significant impact on the performance of quantum circuits. 
In a superconducting device, such as IBM ibm\_quebec \cite{ibm_quantum_quebec}, the median Error per Clifford (ECR) error rate is 6.931e-3, a crucial metric for understanding the fidelity of quantum operations, particularly when implementing error correction protocols. 
The median T1 and T2 time are 284.52 us and 214.31 us respectively. 
These two metrics indicate how long a qubit can retain its quantum state before it loses energy. 
It also has other errors, such as state preparation and measurement (SPAM) errors, crosstalk errors, etc. 

These errors make the execution of quantum programs lose fidelity. 
The impact of these various errors is exacerbated in deep circuits, where the errors accumulate over many operations and execution time, leading to substantial deviations from the expected outcomes. 
As a result, minimizing circuit depth and gate count is not just a matter of efficiency but also essential for reducing the overall error rate and enhancing the reliability of quantum computations.

\begin{figure}
    \centering
    \includegraphics[width=0.85\linewidth]{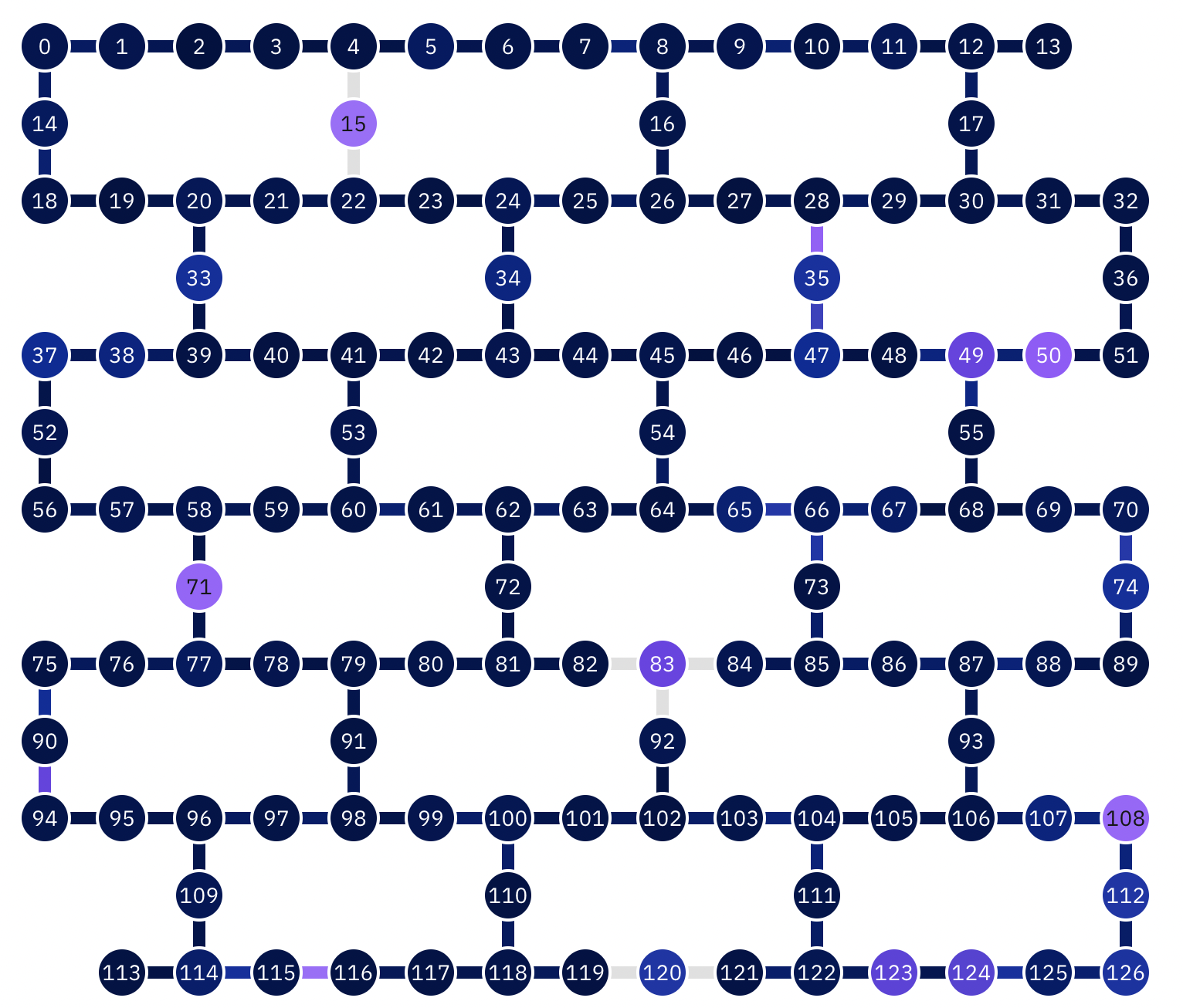}
    \caption{Qubit Connection in IBM\_Quebec.}
    \label{fig:ibm_quebec}
\end{figure}

In addition to the errors in a quantum device, the qubit connection tends to be sparse to avoid crosstalk errors \cite{hua2023crosstalk, Sheldon2016crosstalk, Ding_2020crosstalk}. 
For example, Fig. \ref{fig:ibm_quebec} represents the qubit connection in the ibm\_quebec device with readout error in each qubit and ECR error in each connection. 
Each circle represents a qubit and a link represents where a two-qubit gate could be executed. 
The color indicates the error variability. 
A lighter color indicates a higher error rate. 

Such sparse connectivity requires more SWAP gate insertion in the circuit compilation. 
Those additional SWAP gates increase the gate count, extend the circuit execution, and exacerbate the negative impact of errors in the program execution outcome.

\paragraph{\textbf{Global view in the QAOA compilation.}} 
Qubit mapping is a NP-hard Problem \cite{zhang_asplos21_qubitmapping, li2019tacklingqubitmappingproblem, niu2020hardware,quetschlich2023reducing}. 
The existing generic quantum compilers either leverage SAT solvers \cite{molavi2022qubit,tan2024sat}, which suffer from scalability issues, or use heuristic solutions that do not guarantee the quality of the compilation results~\cite{li2019tackling}. 
The challenge becomes even more significant in the compilation of the QAOA problem, where gate commutativity can be exploited. 
Although, some domain-specific compilers for QAOA program are proposed~\cite{jin2021structured, alam2020circuit, alam2020efficient}, they still face significant drawbacks, such as scalability issues~\cite{tan2024sat, molavi2022qubit} and unexpectedly high compilation costs~\cite{lao20222qan, li2022paulihedral, alam2020circuit}.

However, as the pattern introduced in \cite{jin2023quantum,gao2024linear} for quantum fourier transformation (QFT) circuit compilation in Heavy-hex architecture, we found that there exists a solution with global consideration for the compilation of QAOA.  
QFT circuit has all-to-all interactions with certain dependency constraints. 
The QAOA circuit for the clique input problem graph also has all-to-all interactions. 
We could impose the same gate dependency constraint in QFT to QAOA circuit and reuse the QFT pattern to compile QAOA program. 
Although this gate dependency is unnecessary to QAOA, it gives us an idea of what the upper bound of the compilation cost of QAOA program could be. 

\begin{figure}
        \centering
        \includegraphics[width=0.95\linewidth]{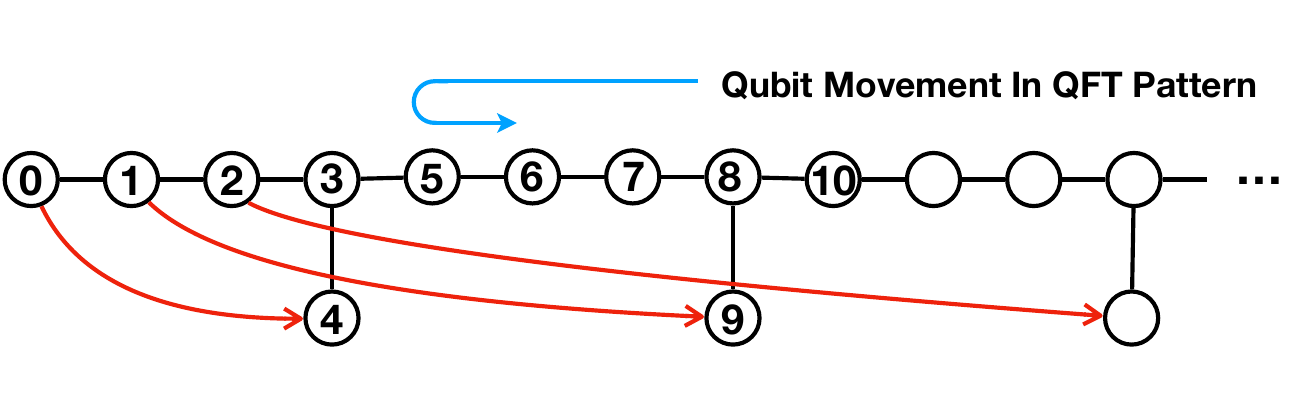}
        \caption{Expanded Heavy-hex Architecture and QFT Pattern.}
        \label{fig:exp_hh}
    \end{figure}

The idea of the QFT pattern in \cite{jin2023quantum,gao2024linear} is stretched in Fig. \ref{fig:exp_hh}. Firstly, it ignores some connections and expands the heavy-hex architecture into a linear nearest-neighbor connected line with certain dangling points. Then, all qubits move to the left and turn around once they hit the boundary. When the qubit with the smallest index moves to the position above the dangling point, it switches with the qubit at the dangling point. For example, qubit 1 will reside at the first dangling point. This pattern gives an upper bound in circuit depth with 5n + O(1), where n is the total number of qubits.

Even though, \cite{jin_2023_qaoa} also introduced a pattern for compiling QAOA in Heavy-hex architecture, the solution consisted of two iterations of linear nearest neighboring (LNN) pattern with some extra interactions between off-path qubits and interactions between path qubits and off-path qubits. However, the upper bound for those extra operations is unclear.

    
\paragraph{\textbf{Adaptation from QFT pattern to generic QAOA.}} There are two types of adaptations we have to do. Adaptation 1: Resolving unnecessary dependencies in QAOA compilation. This adaptation is straightforward. In the QAOA compilation using the QFT pattern, we can directly execute all gates where the corresponding qubits are mapped to connected physical qubits. Some gates in QAOA may be executed earlier than their scheduled execution in the QFT pattern. When we encounter these gates later, we can ignore the operations that have already been scheduled.

Adaptation 2, resolving redundant SWAP gates for sparse input problem graphs. As we discussed above, we can use QFT pattern for compiling the clique problem graph in QAOA. Any graph is the sub-graph of the clique graph. We can simply reuse QFT pattern for a random graph as long as we do not execute gates that the corresponding edges do not appear in the input problem graph. However, the number of SWAP gates and where they are performed is fixed. It is possible that only a few of remaining gates are scheduled in the late stage of the pattern and we have to go through all SWAP gates that are pre-defined in the pattern. Those redundant SWAP gates could be avoided heuristically. For more details about the adaptation, please refer to our next section.

\section{Methodology}
\label{section_methodology}
\subsection{Modified Topology for Repetitive Patterns}
\begin{figure}
    \centering
    \includegraphics[width=\linewidth]{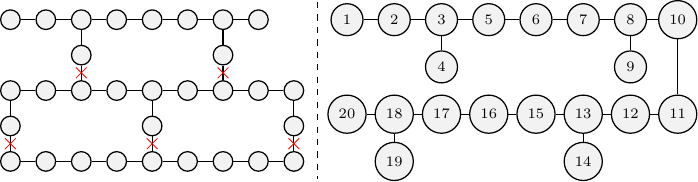}
    \caption{Modified coupling graph of heavy-hex architecture to generate repetitive patterns.}
    \label{fig:coupgraph}
\end{figure}

To take advantages of the topology information during circuit compilation, it is crucial ti identify repeetitive patterns.
However, these patterns can be challenging to recognize directly within a heavy-hex architecture. 
Different from linear nearest neighbour (LNN) and 2D grid, heavy-hex architecture can not be described as the duplication of a group. 
Therefore we process the coupling graph of heavy-hex is through removing some qubit pairs. 
As shown in Fig~\ref{fig:coupgraph}, after removing certain qubit pairs, the final coupling graph resembles the linear structure. 
After the modification, we now have a repetitive pattern that is composed of four qubits in a line and a qubit dangled. 
Also, the qubits' labels in the modified architecture also differ from the original heavy-hex architecture where the label is consecutive in the same line. 
As shown in the right side of Fig~\ref{fig:coupgraph}, the dangling qubit is labeled after the adjacent line qubit. 
This labeling method ensures that previous qubits do not need relabeled when more repetitive patterns are concatenated to the modified architecture.

This modification could offer several advantages. 
Two-qubit entanglement operations require a frequency offset from neighboring qubit transition frequencies to avoid collisions. 
The reduced connectivity in the new architecture results in fewer frequency constraints for two-qubit interactions, potentially enhancing fidelity. 
More importantly, the architecture finds a balanced point between hardware connectivity and software requirements. 
The coupling graph exhibits a clear repetitive pattern while maintaining most connections. 
Additionally, based on the heavy-hex architecture, multiple approaches can be employed to transition to the new architecture. 
For instance, for each dangling qubit, either the upper or lower connection can be removed, providing a broad design space for further optimizations.

     \begin{figure}
        \centering
        \includegraphics[width=\linewidth]{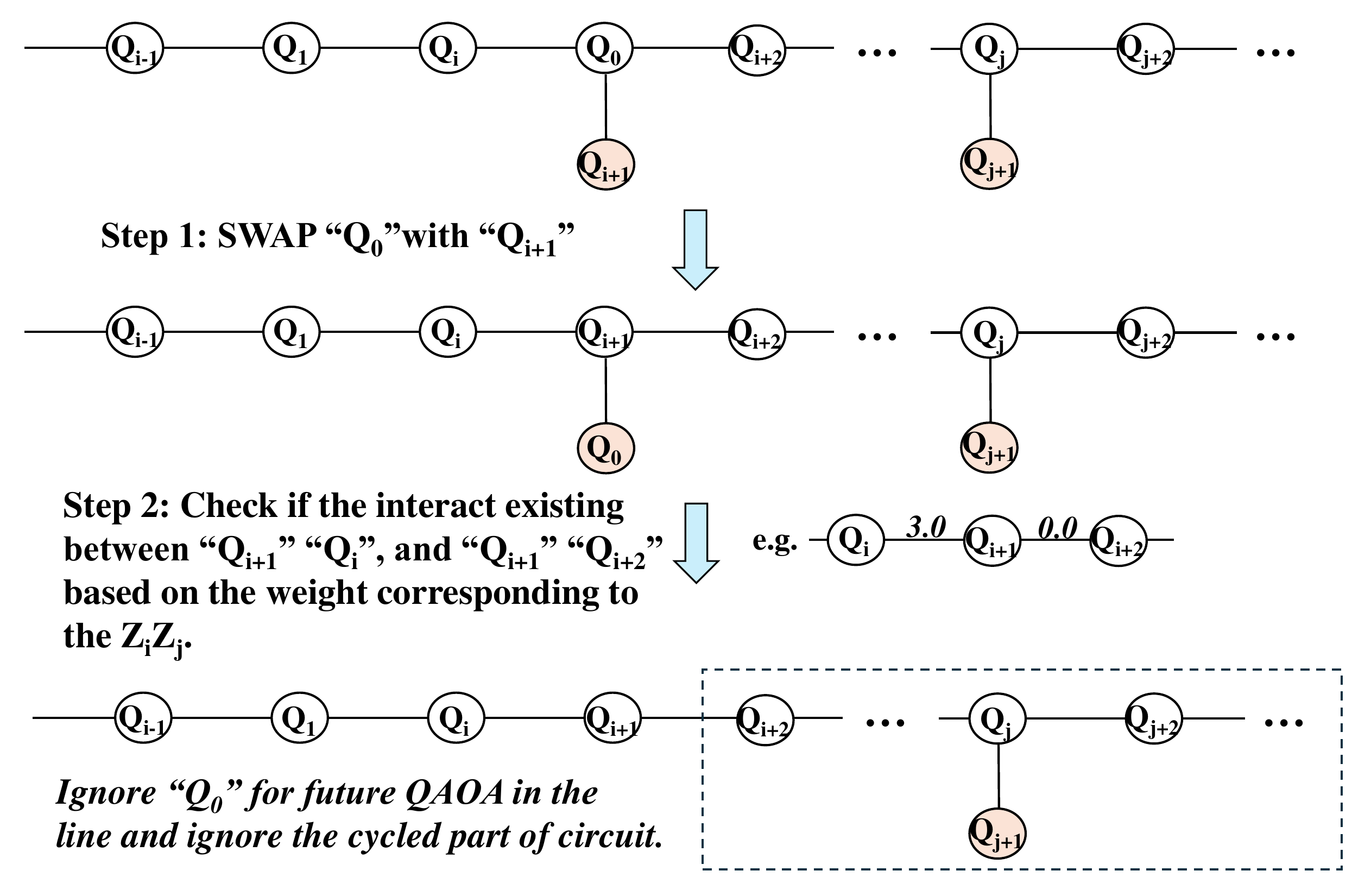}
        \caption{Recursive pattern: reducing one dangling
node from the coupling graph after two steps: one SWAP gate between the qubit above and the dangling point itself, and then check the interaction between pair-formed nearby qubits based on the weight corresponding to the $Z_iZ_j$.}
        \label{fig:qaoaschedule}
    
    \end{figure}

\subsection{Linear Nearest Neighbor}
The linear Nearest neighbor (LNN) architecture is a fundamental model in quantum computing, where qubits are connected in a linear sequence, allowing direct interaction only between adjacent qubits.
This architecture is intuitive and resembles the topologies of many current quantum computers, where qubit connections are sparse and the number of connections grows linearly with the number of qubits.
The linear arrangement in LNN simplifies the overall topology, reducing its complexity and enabling the acceleration of the mapping process with a guaranteed upper bound on the circuit size.
This makes it easier to implement quantum algorithms efficiently on LNN architectures, as opposed to relying on general-purpose compilers.
Recent works~\cite{jin2023quantum} have explored the LNN with Quantum Fourier transform (QFT) algorithm, and we are the first to explore this method on QAOA algorithm. Unlike QFT, where all-to-all qubit connectivity is often beneficial to perform the necessary quantum gates efficiently, QAOA does not necessitate a fully connected qubit architecture, so we need to carefully remove redundant circuit during the compilation. we will introduce details in the following subsections.

\subsection{Linear Structure with Dangling Point}
In a heavy-hex lattice, dangling points are qubits that are connected to the main body of qubits but only have one direct neighbor. 
These qubits are usually evenly distributed on the edge or corner of the qubit lattice, as indicated in Fig.~\ref{fig:ibm_quebec}.
Since dangling points are qubits with limited connectivity, we need to pay special attention to these qubits when we design and implement the mapping process. 
For example, we should consider to place qubits with less connections onto the dangling points to minimize the need of SWAP gates.
We can also use dangling points to store qubits that are not immediately needed but will be used later in the quantum program. The first time qubit $q_0$ is positioned above (connected to) $q_{i+1}$ (the dangling qubit), as illustrated in Figure ~\ref{fig:qaoaschedule}, we execute a SWAP operation on $q_0$ and $q_{i+1}$. This results in the qubits along the main line taking a configuration similar to an intermediate step in the linear QFT process~\cite{jin2023exploiting,gao2024linear}, where the line begins with $q_1$ and ends at $q_{N-1}$, just before $q_1$ swaps with $q_{i+1}$. The process then continues with the standard steps of performing QAOA from $q_1$ to $q_{N-1}$ along the straight line until completion. Importantly, whenever a qubit $q_j$ (where $j > i$) is moved adjacent to $q_0$ in the dangling position for the first time, we pause the algorithm and the algorithm concludes. According to the properties of QAOA, after performing the SWAP operation between $q_0$ and $q_{i+1}$, we need to check the interaction based on the weights in the problem Hamiltonian. For example, in Figure 6, if the weight between $q_{i+1}$ and $q_1$ is 3.0, and the weight between $q_{i+1}$ and $q_{i+2}$ is 0.0, then there is no interaction between $q_{i+1}$ and $q_{i+2}$. In this case, we can safely ignore the redundant circuits following $q_{i+2}$.


\subsection{General Complexity Analysis}
Given our initial labeling, all previous qubits' labels remain unchanged after adding more repetitive groups of qubits and most operations will overlap. Therefore, the time complexity of our algorithms could be inducted through the calculation of extra steps required to add a qubit group. Considering a group of 5 more qubits are added to a finished mapping of $5n$ qubits, the extra operations are composed of two phases: moving $q_{n}$ back towards the final dangling point and moving $q_{5n+4}$ forward towards the first position. It takes 10 steps for moving $q_{n}$ to the dangling points and an extra 15 steps to move $q_{5n+4}$ to the correct place. Therefore, for every 5 qubits, at most 25 steps are required to complete the QAOA mapping, and the time complexity of our mapping algorithm is $5N +O (1)$, ensuring a linear upbound for the mapping process.

\section{Evaluation}
\label{section_evaluation}

\subsection{Experimental Setup}
The experiments in this study were conducted on a laptop equipped with the Apple M1 Pro chip. This system features a 10-core CPU, consisting of 8 performance cores and 2 efficiency cores, supporting 10 concurrent threads with a base clock frequency of 3.2 GHz. The device also includes 16 GB of unified memory. We utilized Qiskit 1.1.0 as the software platform for our experiments. Our results are compared against QAIM (Qubit Allocation and Initial Mapping)~\cite{alam2020circuit} and the structured method proposed by Jin et al.~\cite{jin2021structured}. To create QAOA instances of different graphs, We use the library Networkx in python to generate Erdös-Rényi random graphs and random k-regular graphs. For each type of graphs and each vertex number, we vary graph density to 0.3 and 0.5.

\subsection{Compilation Time}
The Table~\ref{tab:compilation_time} compares the compilation times of various QAOA compilers including QAIM, a Structured Method, and our proposed methods across different qubit counts. 
The results demonstrate that our method consistently outperforms the other methods, showing significantly lower compilation times across all tested qubit counts. 
For instance, at 65 qubits, our method takes only 0.012 seconds, compared to 292 seconds for QAIM and 0.8 seconds for the Structured Method. 
As the qubit count increases, the efficiency of our method remains obvious with a minimal increase in time even for 1025 qubits where it takes 8.68 seconds, while QAIM exceeds 5 hours. 
In summary, our method demonstrates superior performance in terms of compilation time.

\begin{figure*}[htp]
     \centering
     \begin{subfigure}[b]{0.24\textwidth}
        \includegraphics[width=\linewidth]{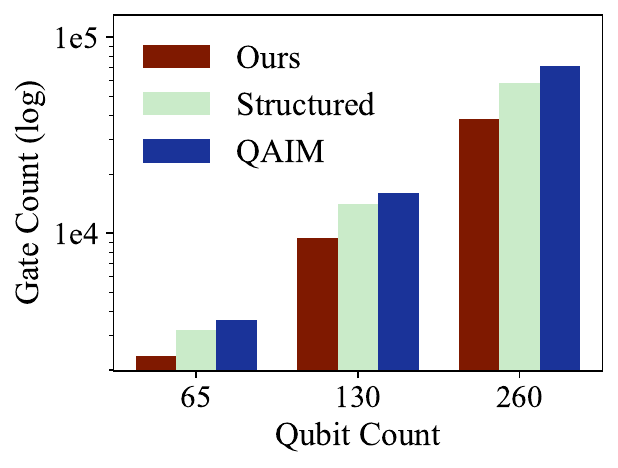}
        \caption{Gate count of random graphs}
        \label{subfig:a}
     \end{subfigure}
     \begin{subfigure}[b]{0.24\textwidth}
        \includegraphics[width=\linewidth]{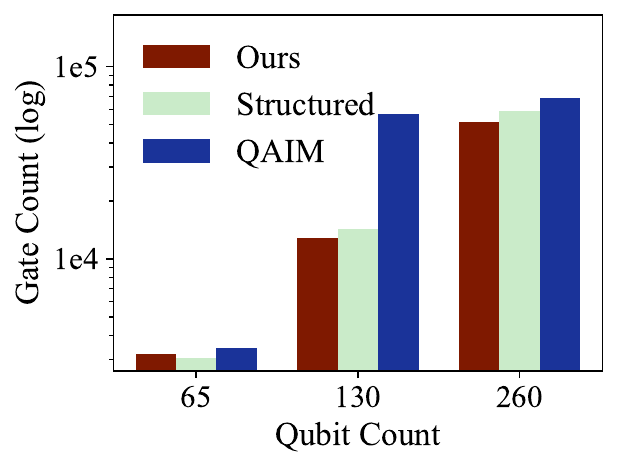}
        \caption{Gate count of regular graphs}
        \label{subfig:b}
     \end{subfigure}
     \begin{subfigure}[b]{0.24\textwidth}
         \includegraphics[width=\linewidth]{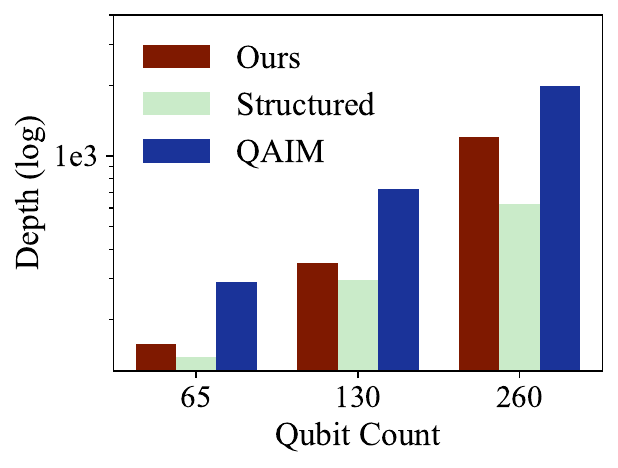}
         \caption{Depth of random graphs}
        \label{subfig:c}
     \end{subfigure}
     \begin{subfigure}[b]{0.24\textwidth}
         \includegraphics[width=\linewidth]{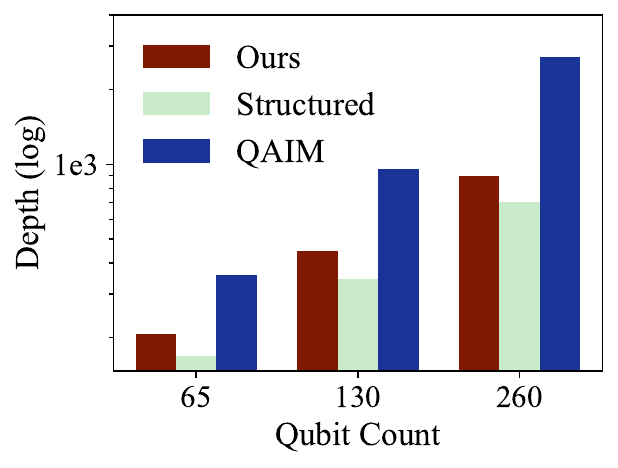}
         \caption{Depth of regular graphs}
        \label{subfig:d}
     \end{subfigure}
    \caption{Comparison of large-scale QAOA circuit compilation results including random graphs and regular graphs.}
    \label{fig:large_scale_results}   
\end{figure*}
\subsection{Results for Small-Scale Circuits}
The results in Table~\ref{tab:circuit_depth_gate_count} compare the performance of three different methods: our method, the structured method, and QAIM. The experiments are conducted with QAOA circuits on different graph instances, evaluating both circuit depth and gate count. These instances are relatively small, and our method achieves significantly shorter compilation times. For these small-scale circuits, our method generally demonstrates comparable circuit depths and gate counts. For example, in the 25-0.3 graph, our method produces 202 gates, compared to 194 for the structured method and 215 for QAIM. The overall trend indicates that our method offers competitive performance with compilation times that are only a fraction of those required by other methods.

\subsection{Results for Large-Scale Circuits}
The graphs in Figure~\ref{fig:large_scale_results} present a comparative analysis of gate count and depth for random and regular graphs with varying qubit counts (65, 130, and 260) using three methods: our method, the structured method, and QAIM. The gate count, as shown on a logarithmic scale in Figure~\ref{subfig:a} and Figure~\ref{subfig:b}, indicates that our method consistently produces the lowest number of gates across all qubit counts for both random and regular graphs. This demonstrates that our method is more efficient in terms of gate count. For circuit depth, as presented in Figure~\ref{subfig:c} and Figure~\ref{subfig:d}, our method consistently outperforms QAIM, showing lower depths for all qubit counts. At the same time, our method's circuit depth is slightly higher and comparable to that of the structured method. Overall, our method is capable of producing the lowest gate counts and competitive circuit depths with very fast compilation times.
 
\begin{table}[h]
    \centering
    \caption{Compilation time comparison of various QAOA compilers. }
    \begin{tabular}{cccc}
    \toprule
         \multirow{2}{*}{Qubit count} & \multicolumn{3}{c}{Time(s)}  \\
         & QAIM & Structured Method & Ours \\
                                        
    \midrule
         {65}      & 292    & 0.8   & 0.012 \\
         125       & 4215    & 6.4  & 0.029 \\
         515       & > 5hrs  & 47.4  & 1.14 \\
         1025      & > 5hrs  & 337   & 8.68 \\
    \bottomrule
    \end{tabular}
    \label{tab:compilation_time}
\end{table}

\begin{table}[h]
    \centering
    \renewcommand*{\arraystretch}{1}
\setlength{\tabcolsep}{1.8pt}
    \caption{Comparison of small-scale QAOA circuit compilation results where graphs are composed of random graphs and regular graphs, e.g., "10-0.3" represents a random graph with 10 nodes and an edge generation probability of 0.3, and "20-6" represents a regular graph with 20 nodes and each node has 6 edges.}
    \begin{tabular}{ccccccc}
    \toprule
        \multirow{2}{*}{Graphs} & \multicolumn{3}{c}{Depth}  & \multicolumn{3}{c}{Gate Count} \\
        & Ours & Structured & QAIM & Ours & Structured & QAIM \\
        \midrule
        10-0.3 &12    &6 & 11 & 19 & 19 &23 \\
        15-0.3 &26  &15 &25 & 59 & 64 & 65 \\
        20-0.3 &33   &22 &33 &95 & 119 &129   \\
        25-0.3 &35   &29 &52 &202 & 194 &215 \\
        10-0.5 &24   &15  &22 & 47 &44 &49 \\
        15-0.5 &29     &17   &32 &89 &87 &95 \\
        20-0.5 & 36   & 24 & 53 &167 & 159 & 190 \\
        25-0.5 & 39 & 34 & 66   & 271  & 259   & 307 \\
        20-6 &28  & 17    & 28 &109 & 107    &113 \\
        20-8 &35  & 22 & 36   & 150  &142    & 155   \\
        25-10 & 45    & 30    & 71    & 247 & 235 &279    \\
        \bottomrule
    \end{tabular}
    \label{tab:circuit_depth_gate_count}
\end{table}

\section{Related Work}
\label{section_relatedwork}
Previous work~\cite{jin2021structured} introduces a method for optimizing QAOA circuits, addressing the limitations of existing compilers in terms of performance and scalability when dealing with large-scale quantum circuits. 
This structured compilation method leverages the regularity of quantum hardware to ensure linear depth on multi-dimensional architectures, significantly outperforming existing approaches in both performance and scalability. 
\cite{alam2020circuit} present four novel methodologies - QAIM (Qubit Allocation and Initial Mapping), IP (Instruction Parallelization), IC (Incremental Compilation), and VIC (Variation-aware Incremental Compilation) — offer significant improvements over existing approaches to optimize the compilation of QAOA circuits, addressing challenges that arise with NISQ devices. 
And~\cite{alam2020efficient} builds on efforts by introducing scalable, compiler-agnostic methodologies that achieve significant reductions in both gate count and circuit depth, improving the overall performance of QAOA circuits. 
Recent work~\cite{jin2023quantum, gao2024linear} proposes circuit mapping solutions tailored for modern multi-dimensional quantum architectures designed for heavy-hex, significantly improving both efficiency and scalability over previous approaches, however, this work is specialized for linear-depth QFT.

\section{Conclusion}
\label{section_conclusion}
In this paper, we propose to take advantages of the patterns in QAOA circuits and the topology of existing quantum computers to optimize the compilation process. Specifically, we unroll the topology of current quantum computers to a linear nearest neighbor (LNN) configuration and implement a tailored mapping algorithm for QAOA circuits which are determined by the graph instances they represent. As a result, we achieve a 30\% reduction in gate count and a 39x acceleration in the compilation process.
\section{Acknowledgment}
We acknowledge the use of IBM Quantum services for this work. 

\clearpage
\bibliographystyle{ACM-Reference-Format}
\bibliography{sample-base}


\begin{thebibliography}{42}


\ifx \showCODEN    \undefined \def \showCODEN     #1{\unskip}     \fi
\ifx \showDOI      \undefined \def \showDOI       #1{#1}\fi
\ifx \showISBNx    \undefined \def \showISBNx     #1{\unskip}     \fi
\ifx \showISBNxiii \undefined \def \showISBNxiii  #1{\unskip}     \fi
\ifx \showISSN     \undefined \def \showISSN      #1{\unskip}     \fi
\ifx \showLCCN     \undefined \def \showLCCN      #1{\unskip}     \fi
\ifx \shownote     \undefined \def \shownote      #1{#1}          \fi
\ifx \showarticletitle \undefined \def \showarticletitle #1{#1}   \fi
\ifx \showURL      \undefined \def \showURL       {\relax}        \fi
\providecommand\bibfield[2]{#2}
\providecommand\bibinfo[2]{#2}
\providecommand\natexlab[1]{#1}
\providecommand\showeprint[2][]{arXiv:#2}

\bibitem[\protect\citeauthoryear{Alam, Ash-Saki, and Ghosh}{Alam et~al\mbox{.}}{2020a}]%
        {alam2020circuit}
\bibfield{author}{\bibinfo{person}{Mahabubul Alam}, \bibinfo{person}{Abdullah Ash-Saki}, {et~al\mbox{.}}} \bibinfo{year}{2020}\natexlab{a}.
\newblock \showarticletitle{Circuit compilation methodologies for quantum approximate optimization algorithm}. In \bibinfo{booktitle}{\emph{2020 53rd Annual IEEE/ACM International Symposium on Microarchitecture (MICRO)}}. IEEE, \bibinfo{pages}{215--228}.
\newblock


\bibitem[\protect\citeauthoryear{Alam, Ash-Saki, and Ghosh}{Alam et~al\mbox{.}}{2020b}]%
        {alam2020efficient}
\bibfield{author}{\bibinfo{person}{Mahabubul Alam}, \bibinfo{person}{Abdullah Ash-Saki}, {et~al\mbox{.}}} \bibinfo{year}{2020}\natexlab{b}.
\newblock \showarticletitle{An efficient circuit compilation flow for quantum approximate optimization algorithm}. In \bibinfo{booktitle}{\emph{2020 57th ACM/IEEE Design Automation Conference (DAC)}}. IEEE, \bibinfo{pages}{1--6}.
\newblock


\bibitem[\protect\citeauthoryear{Aleksandrowicz, Alexander, Barkoutsos, Bello, Ben-Haim, et~al\mbox{.}}{Aleksandrowicz et~al\mbox{.}}{2019}]%
        {aleksandrowicz2019qiskit}
\bibfield{author}{\bibinfo{person}{Gadi Aleksandrowicz}, \bibinfo{person}{Thomas Alexander}, {et~al\mbox{.}}} \bibinfo{year}{2019}\natexlab{}.
\newblock \bibinfo{title}{Qiskit: An open-source framework for quantum computing}.
\newblock
\newblock


\bibitem[\protect\citeauthoryear{Cao, Romero, Olson, Degroote, Johnson, et~al\mbox{.}}{Cao et~al\mbox{.}}{2019}]%
        {cao2019quantum}
\bibfield{author}{\bibinfo{person}{Yudong Cao}, \bibinfo{person}{Jonathan Romero}, {et~al\mbox{.}}} \bibinfo{year}{2019}\natexlab{}.
\newblock \showarticletitle{Quantum chemistry in the age of quantum computing}.
\newblock \bibinfo{journal}{\emph{Chemical reviews}} \bibinfo{volume}{119}, \bibinfo{number}{19} (\bibinfo{year}{2019}), \bibinfo{pages}{10856--10915}.
\newblock


\bibitem[\protect\citeauthoryear{Cheng, Wang, Liang, Shi, Han, et~al\mbox{.}}{Cheng et~al\mbox{.}}{2022}]%
        {cheng2022topgen}
\bibfield{author}{\bibinfo{person}{Jinglei Cheng}, \bibinfo{person}{Hanrui Wang}, {et~al\mbox{.}}} \bibinfo{year}{2022}\natexlab{}.
\newblock \bibinfo{title}{TopGen: Topology-Aware Bottom-Up Generator for Variational Quantum Circuits}.
\newblock
\newblock
\newblock
\shownote{arXiv preprint, \arxiv{2210.08190}.}


\bibitem[\protect\citeauthoryear{Ding, Gokhale, Lin, Rines, Propson, et~al\mbox{.}}{Ding et~al\mbox{.}}{2020}]%
        {Ding_2020crosstalk}
\bibfield{author}{\bibinfo{person}{Yongshan Ding}, \bibinfo{person}{Pranav Gokhale}, {et~al\mbox{.}}} \bibinfo{year}{2020}\natexlab{}.
\newblock \showarticletitle{Systematic Crosstalk Mitigation for Superconducting Qubits via Frequency-Aware Compilation}. In \bibinfo{booktitle}{\emph{2020 53rd Annual IEEE/ACM International Symposium on Microarchitecture (MICRO)}}. \bibinfo{publisher}{IEEE}.
\newblock
\urldef\tempurl%
\url{https://doi.org/10.1109/micro50266.2020.00028}
\showDOI{\tempurl}


\bibitem[\protect\citeauthoryear{Farhi, Goldstone, and Gutmann}{Farhi et~al\mbox{.}}{2014}]%
        {farhi2014quantum}
\bibfield{author}{\bibinfo{person}{Edward Farhi}, \bibinfo{person}{Jeffrey Goldstone}, {et~al\mbox{.}}} \bibinfo{year}{2014}\natexlab{}.
\newblock \showarticletitle{A quantum approximate optimization algorithm}.
\newblock \bibinfo{journal}{\emph{arXiv preprint arXiv:1411.4028}} (\bibinfo{year}{2014}).
\newblock


\bibitem[\protect\citeauthoryear{Gao, Jin, Guo, Chen, and Zhang}{Gao et~al\mbox{.}}{2024}]%
        {gao2024linear}
\bibfield{author}{\bibinfo{person}{Xiangyu Gao}, \bibinfo{person}{Yuwei Jin}, {et~al\mbox{.}}} \bibinfo{year}{2024}\natexlab{}.
\newblock \showarticletitle{Linear Depth QFT over IBM Heavy-hex Architecture}.
\newblock \bibinfo{journal}{\emph{arXiv preprint arXiv:2402.09705}} (\bibinfo{year}{2024}).
\newblock


\bibitem[\protect\citeauthoryear{Guerreschi and Matsuura}{Guerreschi and Matsuura}{2019}]%
        {guerreschi2019qaoa}
\bibfield{author}{\bibinfo{person}{Gian~Giacomo Guerreschi} {and} \bibinfo{person}{Anne~Y. Matsuura}.} \bibinfo{year}{2019}\natexlab{}.
\newblock \showarticletitle{QAOA for Max-Cut requires hundreds of qubits for quantum speed-up}.
\newblock \bibinfo{journal}{\emph{Scientific reports}} \bibinfo{volume}{9}, \bibinfo{number}{1} (\bibinfo{year}{2019}), \bibinfo{pages}{1--7}.
\newblock


\bibitem[\protect\citeauthoryear{Hua, Jin, Li, Liu, Wang, et~al\mbox{.}}{Hua et~al\mbox{.}}{2023}]%
        {hua2023crosstalk}
\bibfield{author}{\bibinfo{person}{Fei Hua}, \bibinfo{person}{Yuwei Jin}, {et~al\mbox{.}}} \bibinfo{year}{2023}\natexlab{}.
\newblock \bibinfo{title}{A Synergistic Compilation Workflow for Tackling Crosstalk in Quantum Machines}.
\newblock
\newblock
\showeprint[arxiv]{quant-ph/2207.05751}
\urldef\tempurl%
\url{https://arxiv.org/abs/2207.05751}
\showURL{%
\tempurl}


\bibitem[\protect\citeauthoryear{{IBM Quantum}}{{IBM Quantum}}{2024}]%
        {ibm_quantum_quebec}
\bibfield{author}{\bibinfo{person}{{IBM Quantum}}.} \bibinfo{year}{2024}\natexlab{}.
\newblock \bibinfo{title}{IBM Quantum Resources}.
\newblock \bibinfo{howpublished}{\url{https://quantum.ibm.com/services/resources?system=ibm_quebec}}.
\newblock
\newblock
\shownote{Accessed: 2024-07-22.}


\bibitem[\protect\citeauthoryear{Jin, Gao, Guo, Chen, Hua, et~al\mbox{.}}{Jin et~al\mbox{.}}{2023a}]%
        {jin2023quantum}
\bibfield{author}{\bibinfo{person}{Yuwei Jin}, \bibinfo{person}{Xiangyu Gao}, {et~al\mbox{.}}} \bibinfo{year}{2023}\natexlab{a}.
\newblock \showarticletitle{Quantum Fourier Transformation Circuits Compilation}.
\newblock \bibinfo{journal}{\emph{arXiv preprint arXiv:2312.16114}} (\bibinfo{year}{2023}).
\newblock


\bibitem[\protect\citeauthoryear{Jin, Hua, Chen, Hayes, Zhang, et~al\mbox{.}}{Jin et~al\mbox{.}}{2023b}]%
        {jin2023exploiting}
\bibfield{author}{\bibinfo{person}{Yuwei Jin}, \bibinfo{person}{Fei Hua}, {et~al\mbox{.}}} \bibinfo{year}{2023}\natexlab{b}.
\newblock \showarticletitle{Exploiting the Regular Structure of Modern Quantum Architectures for Compiling and Optimizing Programs with Permutable Operators}. In \bibinfo{booktitle}{\emph{Proceedings of the 28th ACM International Conference on Architectural Support for Programming Languages and Operating Systems, Volume 4}}. \bibinfo{pages}{108--124}.
\newblock


\bibitem[\protect\citeauthoryear{Jin, Hua, Chen, Hayes, Zhang, et~al\mbox{.}}{Jin et~al\mbox{.}}{2024a}]%
        {jin_2023_qaoa}
\bibfield{author}{\bibinfo{person}{Yuwei Jin}, \bibinfo{person}{Fei Hua}, {et~al\mbox{.}}} \bibinfo{year}{2024}\natexlab{a}.
\newblock \showarticletitle{Exploiting the Regular Structure of Modern Quantum Architectures for Compiling and Optimizing Programs with Permutable Operators}. In \bibinfo{booktitle}{\emph{Proceedings of the 28th ACM International Conference on Architectural Support for Programming Languages and Operating Systems, Volume 4}} \emph{(\bibinfo{series}{ASPLOS '23})}. \bibinfo{publisher}{Association for Computing Machinery}, \bibinfo{address}{New York, NY, USA}, \bibinfo{pages}{108–124}.
\newblock
\showISBNx{9798400703942}
\urldef\tempurl%
\url{https://doi.org/10.1145/3623278.3624751}
\showDOI{\tempurl}


\bibitem[\protect\citeauthoryear{Jin, Li, Hua, Hao, Zhou, et~al\mbox{.}}{Jin et~al\mbox{.}}{2024b}]%
        {jin2024tetris}
\bibfield{author}{\bibinfo{person}{Yuwei Jin}, \bibinfo{person}{Zirui Li}, {et~al\mbox{.}}} \bibinfo{year}{2024}\natexlab{b}.
\newblock \bibinfo{title}{Tetris: A Compilation Framework for VQA Applications in Quantum Computing}.
\newblock
\newblock
\showeprint[arxiv]{quant-ph/2309.01905}
\urldef\tempurl%
\url{https://arxiv.org/abs/2309.01905}
\showURL{%
\tempurl}


\bibitem[\protect\citeauthoryear{Jin, Luo, Fong, Chen, Hayes, et~al\mbox{.}}{Jin et~al\mbox{.}}{2021}]%
        {jin2021structured}
\bibfield{author}{\bibinfo{person}{Yuwei Jin}, \bibinfo{person}{Jason Luo}, {et~al\mbox{.}}} \bibinfo{year}{2021}\natexlab{}.
\newblock \showarticletitle{A structured method for compilation of qaoa circuits in quantum computing}.
\newblock \bibinfo{journal}{\emph{arXiv preprint arXiv:2112.06143}} (\bibinfo{year}{2021}).
\newblock


\bibitem[\protect\citeauthoryear{Lao and Browne}{Lao and Browne}{2022}]%
        {lao20222qan}
\bibfield{author}{\bibinfo{person}{Lingling Lao} {and} \bibinfo{person}{Dan~E Browne}.} \bibinfo{year}{2022}\natexlab{}.
\newblock \showarticletitle{2qan: A quantum compiler for 2-local qubit hamiltonian simulation algorithms}. In \bibinfo{booktitle}{\emph{Proceedings of the 49th Annual International Symposium on Computer Architecture}}. \bibinfo{pages}{351--365}.
\newblock


\bibitem[\protect\citeauthoryear{Li, Ding, and Xie}{Li et~al\mbox{.}}{2019a}]%
        {li2019tacklingqubitmappingproblem}
\bibfield{author}{\bibinfo{person}{Gushu Li}, \bibinfo{person}{Yufei Ding}, {et~al\mbox{.}}} \bibinfo{year}{2019}\natexlab{a}.
\newblock \bibinfo{title}{Tackling the Qubit Mapping Problem for NISQ-Era Quantum Devices}.
\newblock
\newblock
\showeprint[arxiv]{cs.ET/1809.02573}
\urldef\tempurl%
\url{https://arxiv.org/abs/1809.02573}
\showURL{%
\tempurl}


\bibitem[\protect\citeauthoryear{Li, Ding, and Xie}{Li et~al\mbox{.}}{2019b}]%
        {li2019tackling}
\bibfield{author}{\bibinfo{person}{Gushu Li}, \bibinfo{person}{Yufei Ding}, {et~al\mbox{.}}} \bibinfo{year}{2019}\natexlab{b}.
\newblock \showarticletitle{Tackling the qubit mapping problem for NISQ-era quantum devices}. In \bibinfo{booktitle}{\emph{Proceedings of the twenty-fourth international conference on architectural support for programming languages and operating systems}}. \bibinfo{pages}{1001--1014}.
\newblock


\bibitem[\protect\citeauthoryear{Li, Wu, Shi, Javadi-Abhari, Ding, et~al\mbox{.}}{Li et~al\mbox{.}}{2022}]%
        {li2022paulihedral}
\bibfield{author}{\bibinfo{person}{Gushu Li}, \bibinfo{person}{Anbang Wu}, {et~al\mbox{.}}} \bibinfo{year}{2022}\natexlab{}.
\newblock \showarticletitle{Paulihedral: a generalized block-wise compiler optimization framework for quantum simulation kernels}. In \bibinfo{booktitle}{\emph{Proceedings of the 27th ACM International Conference on Architectural Support for Programming Languages and Operating Systems}}. \bibinfo{pages}{554--569}.
\newblock


\bibitem[\protect\citeauthoryear{Li, Tian, Liu, Peng, and Jiao}{Li et~al\mbox{.}}{2020}]%
        {li2020quantum}
\bibfield{author}{\bibinfo{person}{Yangyang Li}, \bibinfo{person}{Mengzhuo Tian}, {et~al\mbox{.}}} \bibinfo{year}{2020}\natexlab{}.
\newblock \showarticletitle{Quantum optimization and quantum learning: A survey}.
\newblock \bibinfo{journal}{\emph{Ieee Access}}  \bibinfo{volume}{8} (\bibinfo{year}{2020}), \bibinfo{pages}{23568--23593}.
\newblock


\bibitem[\protect\citeauthoryear{Liang, Cheng, Ren, Wang, Hua, et~al\mbox{.}}{Liang et~al\mbox{.}}{2024}]%
        {liang2024napa}
\bibfield{author}{\bibinfo{person}{Zhiding Liang}, \bibinfo{person}{Jinglei Cheng}, {et~al\mbox{.}}} \bibinfo{year}{2024}\natexlab{}.
\newblock \showarticletitle{NAPA: intermediate-level variational native-pulse ansatz for variational quantum algorithms}.
\newblock \bibinfo{journal}{\emph{IEEE Transactions on Computer-Aided Design of Integrated Circuits and Systems}} (\bibinfo{year}{2024}).
\newblock


\bibitem[\protect\citeauthoryear{Liang, Song, Cheng, He, Liu, et~al\mbox{.}}{Liang et~al\mbox{.}}{2023}]%
        {liang2023hybrid}
\bibfield{author}{\bibinfo{person}{Zhiding Liang}, \bibinfo{person}{Zhixin Song}, {et~al\mbox{.}}} \bibinfo{year}{2023}\natexlab{}.
\newblock \showarticletitle{Hybrid gate-pulse model for variational quantum algorithms}. In \bibinfo{booktitle}{\emph{2023 60th ACM/IEEE Design Automation Conference (DAC)}}. IEEE, \bibinfo{pages}{1--6}.
\newblock


\bibitem[\protect\citeauthoryear{Liang, Wang, Cheng, Ding, Ren, et~al\mbox{.}}{Liang et~al\mbox{.}}{2022}]%
        {liang2022variational}
\bibfield{author}{\bibinfo{person}{Zhiding Liang}, \bibinfo{person}{Hanrui Wang}, {et~al\mbox{.}}} \bibinfo{year}{2022}\natexlab{}.
\newblock \showarticletitle{Variational quantum pulse learning}. In \bibinfo{booktitle}{\emph{2022 IEEE International Conference on Quantum Computing and Engineering (QCE)}}. IEEE, \bibinfo{pages}{556--565}.
\newblock


\bibitem[\protect\citeauthoryear{Molavi, Xu, Diges, Pick, Tannu, et~al\mbox{.}}{Molavi et~al\mbox{.}}{2022}]%
        {molavi2022qubit}
\bibfield{author}{\bibinfo{person}{Abtin Molavi}, \bibinfo{person}{Amanda Xu}, {et~al\mbox{.}}} \bibinfo{year}{2022}\natexlab{}.
\newblock \showarticletitle{Qubit mapping and routing via MaxSAT}. In \bibinfo{booktitle}{\emph{2022 55th IEEE/ACM international symposium on Microarchitecture (MICRO)}}. IEEE, \bibinfo{pages}{1078--1091}.
\newblock


\bibitem[\protect\citeauthoryear{Niu, Suau, Staffelbach, and Todri-Sanial}{Niu et~al\mbox{.}}{2020}]%
        {niu2020hardware}
\bibfield{author}{\bibinfo{person}{Siyuan Niu}, \bibinfo{person}{Adrien Suau}, {et~al\mbox{.}}} \bibinfo{year}{2020}\natexlab{}.
\newblock \showarticletitle{A hardware-aware heuristic for the qubit mapping problem in the nisq era}.
\newblock \bibinfo{journal}{\emph{IEEE Transactions on Quantum Engineering}}  \bibinfo{volume}{1} (\bibinfo{year}{2020}), \bibinfo{pages}{1--14}.
\newblock


\bibitem[\protect\citeauthoryear{Peruzzo, McClean, Shadbolt, Yung, Zhou, et~al\mbox{.}}{Peruzzo et~al\mbox{.}}{2014}]%
        {peruzzo+:nature14}
\bibfield{author}{\bibinfo{person}{Alberto Peruzzo}, \bibinfo{person}{Jarrod McClean}, {et~al\mbox{.}}} \bibinfo{year}{2014}\natexlab{}.
\newblock \showarticletitle{A variational eigenvalue solver on a photonic quantum processor}. In \bibinfo{booktitle}{\emph{Nature Communications}}, Vol.~\bibinfo{volume}{5}. \bibinfo{pages}{4213}.
\newblock
\showISBNx{2041-1723}
\urldef\tempurl%
\url{https://doi.org/10.1145/237814.237866}
\showDOI{\tempurl}


\bibitem[\protect\citeauthoryear{Phalak and Ghosh}{Phalak and Ghosh}{2024}]%
        {phalak2024non}
\bibfield{author}{\bibinfo{person}{Koustubh Phalak} {and} \bibinfo{person}{Swaroop Ghosh}.} \bibinfo{year}{2024}\natexlab{}.
\newblock \showarticletitle{Non-parametric Greedy Optimization of Parametric Quantum Circuits}. In \bibinfo{booktitle}{\emph{2024 25th International Symposium on Quality Electronic Design (ISQED)}}. IEEE, \bibinfo{pages}{1--7}.
\newblock


\bibitem[\protect\citeauthoryear{Preskill}{Preskill}{2018}]%
        {Preskill2018NISQ}
\bibfield{author}{\bibinfo{person}{John Preskill}.} \bibinfo{year}{2018}\natexlab{}.
\newblock \showarticletitle{Quantum Computing in the {NISQ} era and beyond}.
\newblock \bibinfo{journal}{\emph{{Quantum}}}  \bibinfo{volume}{2} (\bibinfo{year}{2018}), \bibinfo{pages}{79}.
\newblock
\showISSN{2521-327X}


\bibitem[\protect\citeauthoryear{Quetschlich, Burgholzer, and Wille}{Quetschlich et~al\mbox{.}}{2023}]%
        {quetschlich2023reducing}
\bibfield{author}{\bibinfo{person}{Nils Quetschlich}, \bibinfo{person}{Lukas Burgholzer}, {et~al\mbox{.}}} \bibinfo{year}{2023}\natexlab{}.
\newblock \showarticletitle{Reducing the compilation time of quantum circuits using pre-compilation on the gate level}. In \bibinfo{booktitle}{\emph{2023 IEEE International Conference on Quantum Computing and Engineering (QCE)}}, Vol.~\bibinfo{volume}{1}. IEEE, \bibinfo{pages}{757--767}.
\newblock


\bibitem[\protect\citeauthoryear{Shaydulin, Li, Chakrabarti, DeCross, Herman, et~al\mbox{.}}{Shaydulin et~al\mbox{.}}{2024}]%
        {shaydulin2024evidence}
\bibfield{author}{\bibinfo{person}{Ruslan Shaydulin}, \bibinfo{person}{Changhao Li}, {et~al\mbox{.}}} \bibinfo{year}{2024}\natexlab{}.
\newblock \showarticletitle{Evidence of scaling advantage for the quantum approximate optimization algorithm on a classically intractable problem}.
\newblock \bibinfo{journal}{\emph{Science Advances}} \bibinfo{volume}{10}, \bibinfo{number}{22} (\bibinfo{year}{2024}), \bibinfo{pages}{eadm6761}.
\newblock


\bibitem[\protect\citeauthoryear{Sheldon, Magesan, Chow, and Gambetta}{Sheldon et~al\mbox{.}}{2016}]%
        {Sheldon2016crosstalk}
\bibfield{author}{\bibinfo{person}{Sarah Sheldon}, \bibinfo{person}{Easwar Magesan}, {et~al\mbox{.}}} \bibinfo{year}{2016}\natexlab{}.
\newblock \showarticletitle{Procedure for systematically tuning up cross-talk in the cross-resonance gate}.
\newblock \bibinfo{journal}{\emph{Physical Review A}} \bibinfo{volume}{93}, \bibinfo{number}{6} (\bibinfo{date}{June} \bibinfo{year}{2016}).
\newblock
\showISSN{2469-9934}
\urldef\tempurl%
\url{https://doi.org/10.1103/physreva.93.060302}
\showDOI{\tempurl}


\bibitem[\protect\citeauthoryear{Shor}{Shor}{1999}]%
        {shor1999polynomial}
\bibfield{author}{\bibinfo{person}{Peter~W. Shor}.} \bibinfo{year}{1999}\natexlab{}.
\newblock \showarticletitle{Polynomial-time algorithms for prime factorization and discrete logarithms on a quantum computer}.
\newblock \bibinfo{journal}{\emph{SIAM review}} \bibinfo{volume}{41}, \bibinfo{number}{2} (\bibinfo{year}{1999}), \bibinfo{pages}{303--332}.
\newblock


\bibitem[\protect\citeauthoryear{Sivarajah, Dilkes, Cowtan, Simmons, Edgington, et~al\mbox{.}}{Sivarajah et~al\mbox{.}}{2020}]%
        {Sivarajah_TKET_A_Retargetable_2020}
\bibfield{author}{\bibinfo{person}{Seyon Sivarajah}, \bibinfo{person}{Silas Dilkes}, {et~al\mbox{.}}} \bibinfo{year}{2020}\natexlab{}.
\newblock \showarticletitle{{TKET: A Retargetable Compiler for NISQ devices}}.
\newblock \bibinfo{journal}{\emph{Quantum Science and Technology}}  \bibinfo{volume}{6} (\bibinfo{date}{Nov.} \bibinfo{year}{2020}).
\newblock
\urldef\tempurl%
\url{https://doi.org/10.1088/2058-9565/ab8e92}
\showDOI{\tempurl}


\bibitem[\protect\citeauthoryear{Tan, Niu, and Gidney}{Tan et~al\mbox{.}}{2024}]%
        {tan2024sat}
\bibfield{author}{\bibinfo{person}{Daniel~Bochen Tan}, \bibinfo{person}{Murphy~Yuezhen Niu}, {et~al\mbox{.}}} \bibinfo{year}{2024}\natexlab{}.
\newblock \showarticletitle{A SAT Scalpel for Lattice Surgery: Representation and Synthesis of Subroutines for Surface-Code Fault-Tolerant Quantum Computing}.
\newblock \bibinfo{journal}{\emph{arXiv preprint arXiv:2404.18369}} (\bibinfo{year}{2024}).
\newblock


\bibitem[\protect\citeauthoryear{Wang, Ding, Gu, Lin, Pan, et~al\mbox{.}}{Wang et~al\mbox{.}}{2022a}]%
        {wang2022quantumnas}
\bibfield{author}{\bibinfo{person}{Hanrui Wang}, \bibinfo{person}{Yongshan Ding}, {et~al\mbox{.}}} \bibinfo{year}{2022}\natexlab{a}.
\newblock \showarticletitle{Quantumnas: Noise-adaptive search for robust quantum circuits}. In \bibinfo{booktitle}{\emph{HPCA2022}}. IEEE, \bibinfo{publisher}{IEEE}.
\newblock


\bibitem[\protect\citeauthoryear{Wang, Li, Gu, Ding, Pan, et~al\mbox{.}}{Wang et~al\mbox{.}}{2022b}]%
        {wang2022qoc}
\bibfield{author}{\bibinfo{person}{Hanrui Wang}, \bibinfo{person}{Zirui Li}, {et~al\mbox{.}}} \bibinfo{year}{2022}\natexlab{b}.
\newblock \showarticletitle{QOC: quantum on-chip training with parameter shift and gradient pruning}. In \bibinfo{booktitle}{\emph{Proceedings of the 59th ACM/IEEE Design Automation Conference}}. \bibinfo{publisher}{IEEE}, \bibinfo{address}{San Francisco, USA}, \bibinfo{pages}{655--660}.
\newblock


\bibitem[\protect\citeauthoryear{Younis, Iancu, Lavrijsen, Davis, Smith, et~al\mbox{.}}{Younis et~al\mbox{.}}{2021}]%
        {osti_1785933}
\bibfield{author}{\bibinfo{person}{Ed Younis}, \bibinfo{person}{Costin~C Iancu}, {et~al\mbox{.}}} \bibinfo{year}{2021}\natexlab{}.
\newblock \bibinfo{title}{Berkeley Quantum Synthesis Toolkit (BQSKit) v1}.
\newblock
\newblock
\urldef\tempurl%
\url{https://doi.org/10.11578/dc.20210603.2}
\showDOI{\tempurl}


\bibitem[\protect\citeauthoryear{Zhang, Hayes, Qiu, Jin, Chen, et~al\mbox{.}}{Zhang et~al\mbox{.}}{2021}]%
        {zhang_asplos21_qubitmapping}
\bibfield{author}{\bibinfo{person}{Chi Zhang}, \bibinfo{person}{Ari~B. Hayes}, {et~al\mbox{.}}} \bibinfo{year}{2021}\natexlab{}.
\newblock \showarticletitle{Time-optimal Qubit mapping}. In \bibinfo{booktitle}{\emph{Proceedings of the 26th ACM International Conference on Architectural Support for Programming Languages and Operating Systems}} \emph{(\bibinfo{series}{ASPLOS '21})}. \bibinfo{publisher}{Association for Computing Machinery}, \bibinfo{address}{New York, NY, USA}, \bibinfo{pages}{360–374}.
\newblock
\showISBNx{9781450383172}
\urldef\tempurl%
\url{https://doi.org/10.1145/3445814.3446706}
\showDOI{\tempurl}


\bibitem[\protect\citeauthoryear{Zhang, Paredes, Sundar, Quiroga, Kyrillidis, et~al\mbox{.}}{Zhang et~al\mbox{.}}{2024}]%
        {zhang2024groverqaoa3}
\bibfield{author}{\bibinfo{person}{Zewen Zhang}, \bibinfo{person}{Roger Paredes}, {et~al\mbox{.}}} \bibinfo{year}{2024}\natexlab{}.
\newblock \bibinfo{title}{Grover-QAOA for 3-SAT: Quadratic Speedup, Fair-Sampling, and Parameter Clustering}.
\newblock
\newblock
\showeprint[arxiv]{quant-ph/2402.02585}
\urldef\tempurl%
\url{https://arxiv.org/abs/2402.02585}
\showURL{%
\tempurl}


\bibitem[\protect\citeauthoryear{Zhu, Zhang, Sundar, Green, Alderete, et~al\mbox{.}}{Zhu et~al\mbox{.}}{2022}]%
        {zhu_2023_qaoa}
\bibfield{author}{\bibinfo{person}{Yingyue Zhu}, \bibinfo{person}{Zewen Zhang}, {et~al\mbox{.}}} \bibinfo{year}{2022}\natexlab{}.
\newblock \showarticletitle{Multi-round QAOA and advanced mixers on a trapped-ion quantum computer}.
\newblock \bibinfo{journal}{\emph{Quantum Science and Technology}} \bibinfo{volume}{8}, \bibinfo{number}{1} (\bibinfo{date}{nov} \bibinfo{year}{2022}), \bibinfo{pages}{015007}.
\newblock
\urldef\tempurl%
\url{https://doi.org/10.1088/2058-9565/ac91ef}
\showDOI{\tempurl}


\bibitem[\protect\citeauthoryear{Zhuang, Cunningham, and Guan}{Zhuang et~al\mbox{.}}{2024}]%
        {zhuang2024improving}
\bibfield{author}{\bibinfo{person}{Jun Zhuang}, \bibinfo{person}{Jack Cunningham}, {et~al\mbox{.}}} \bibinfo{year}{2024}\natexlab{}.
\newblock \showarticletitle{Improving Trainability of Variational Quantum Circuits via Regularization Strategies}.
\newblock \bibinfo{journal}{\emph{arXiv preprint arXiv:2405.01606}} (\bibinfo{year}{2024}).
\newblock


\end{thebibliography}










\end{document}